\documentclass[conference]{IEEEtran}
\IEEEoverridecommandlockouts
\usepackage{float}
\usepackage{caption}
\usepackage{subcaption}
\usepackage{cite}
\usepackage{amsmath,amssymb,amsfonts}
\usepackage[noend]{algpseudocode}
\usepackage{algorithm}
\usepackage{graphicx}
\usepackage{subcaption}
\usepackage{textcomp}
\usepackage{xcolor}
\usepackage{hyperref}
\usepackage{amsmath}
\usepackage{cancel}

\def\BibTeX{{\rm B\kern-.05em{\sc i\kern-.025em b}\kern-.08em
    T\kern-.1667em\lower.7ex\hbox{E}\kern-.125emX}}

\renewcommand{\baselinestretch}{0.966}

\begin{document}

\title{Incremental Bounded Model Checking of \\ Artificial Neural Networks in CUDA}

\author{\IEEEauthorblockN{Luiz H. Sena$^{1}$, Iury V. Bessa$^{1}$, Mikhail R. Gadelha$^2$, Lucas C. Cordeiro$^{3}$, and Edjard Mota$^{1}$}
\IEEEauthorblockA{$^{1}$Federal University of Amazonas, Manaus, Brazil \\
$^{2}$SIDIA Instituto de Ci\^encia e Tecnologia, Manaus, Brazil \\
$^{3}$University of Manchester, Manchester, United Kingdom}}

\maketitle

\begin{abstract}
Artificial Neural networks (ANNs) are powerful computing systems employed for various applications
due to their versatility to generalize and to respond to unexpected inputs/patterns.
However, implementations of ANNs for safety-critical systems might lead to failures,
which are hardly predicted in the design phase since ANNs are highly parallel and their parameters are hardly interpretable.
Here we develop and evaluate a novel symbolic software verification framework based on incremental bounded model checking (BMC)
to check for adversarial cases and coverage methods in multi-layer perceptron (MLP).
In particular, we further develop the efficient SMT-based Context-Bounded Model Checker for
Graphical Processing Units (ESBMC-GPU) in order to ensure the reliability of certain safety properties
in which safety-critical systems can fail and make incorrect decisions, thereby leading to unwanted material damage or even put lives in danger.
This paper marks the first symbolic verification framework to reason over ANNs implemented in CUDA.
Our experimental results show that our approach implemented in ESBMC-GPU can successfully
verify safety properties and covering methods in ANNs and correctly generate $28$ adversarial cases in MLPs.
\end{abstract}


\section{Introduction}
\label{sec:intro}

Artificial Neural Networks (ANNs) have been used to perform various tasks in a safety-critical domain (e.g.,
medical diagnosis~\cite{amato2013artificial} and self-driving autonomous cars~\cite{bojarski2016end}),
where failures such as misclassifications can put lives in danger.
In particular, small noises and adversarial perturbations can lead ANNs to misclassify simple patterns and cause
serious damage in safety-critical systems, e.g., a red traffic light misclassified as green.
If these systems are formally verified w.r.t. safety properties,
then guarantees could be provided to ensure reliability and avoid errors in safety-critical ANNs.
Given the current literature, there exists a gap between implementations of ANNs and safety guarantees.
The application of formal methods to ensure these safety properties are not properly consolidated yet, e.g.,
the application of symbolic (software) model checking is still its infancy to solve this problem~\cite{abs-1807-10439}.

There exist various attempts to apply verification and validation techniques to ensure safety in ANNs.
In recent days, testing~\cite{sun2018testing}, training~\cite{zheng2016improving}, and validation~\cite{krogh1995neural}
have achieved impressive results. Sun et al.~\cite{sun2018testing} show that concolic testing provides reasonable results for
generating adversarial cases; the authors describe how unsafe can ANNs be but they do not address
actual implementations of ANN to provide formal safety guarantees.
Zheng et al.~\cite{zheng2016improving} show how training techniques
can bring noise robustness to the trained ANN, but no formal guarantees from a safety perspective are provided.
Krogh and Vedelsby~\cite{krogh1995neural} describe validations techniques during the training phase and
how the training phase can be optimized by exploring the dataset and improving generalization capability.
These studies, however, do not provide safety guarantees for the implementation of ANNs.
Despite positive recent results obtained from these techniques, symbolic software model checking
can be seen as an alternative to ensure the safety of actual implementations of ANN.

In recent years, researchers proposed the verification of ANNs using formal methods~\cite{huang2017safety, sun2018testing, katz2017reluplex}.
Huang et al.~\cite{huang2017safety} describe a formal verification framework based on Satisfiability Modulo Theories (SMT);
they focus on verifying safety classification of images with real cases perturbations such as scratches, changes in camera angle,
or lighting conditions. Sun et al.~\cite{sun2018testing} describe and evaluate concolic testing for ANNs;
their test criteria measure how adversarial an input is w.r.t. other images. These test criteria are used by an algorithm
based on linear programming, which generates test cases by perturbing a given system until adversarial cases are obtained.
Katz et al.~\cite{katz2017reluplex} describe a formal verification technique for checking properties of deep neural networks (DNNs);
their technique extended the simplex method to handle ReLU (Rectified Linear Unit) activation function, which
is evaluated on the airborne collision avoidance system for unmanned aircraft (ACAS Xu).
However, for image classification, the number of layers and neurons is still a problem for formal verification approaches
mainly due to a large number of mathematical computations required by the underlying ANN.

Here we describe and evaluate a novel symbolic software verification framework based on incremental bounded model checking (BMC)
to check covering methods and adversarial cases in ANNs using the efficient SMT-based Context-Bounded Model Checker for Graphical
Processing Units (ESBMC-GPU)~\cite{PereiraASMMFC17,monteiro2018esbmc}.
Our approach, implemented on top of ESBMC-GPU, verifies the actual implementation of ANNs using CUDA and
its deep learning primitives cuDNN~\cite{chetlur2014cudnn} and cuBLAS~\cite{nvidia2008cublas}.
Our work makes two major contributions. First, we formally verify covering methods, as described by Sun et al.~\cite{sun2018testing},
to measure how adversarial are two images w.r.t. the ANN neurons using incremental BMC.
Second, we formally check adversarial cases using instances of ANNs,
where only weights and bias descriptions are needed for our verification process.
In both checks, we use incremental BMC to iteratively verify the implementation of
ANNs for each unwind bound indefinitely (i.e., until it exhausts the time or memory limits).
Intuitively, we aim to either find a counterexample with up to \textit{k} loop unwinding or to fully
unwind all loops in the actual implementations of ANNs so we can provide a correct result.
Our incremental verification algorithm relies on a symbolic execution engine to
increasingly unwind the loop after each iteration of the algorithm.


\section{Preliminaries}
\label{sec:preliminaries}

\subsection{Artificial Neural Networks (ANNs)}
\label{ssec:ann}

Artificial Neural Networks (ANNs) are efficient models
for statistical pattern recognition, which makes them a suitable
paradigm for learning tasks~\cite{bishop2006PRML}.
Learning in ANNs is based on modifying the {\it synaptic} weights of their interconnected units, which will fit
their values according to a given set of labeled examples or tasks, during the training phase. A training
algorithm uses each input example to converge these weights to specific results.
In most applications areas of ANNs, this may solve classification problems and some specific
generalization for adversarial situations~\cite{hagan1996neural}.


There exist various algorithms to train ANNs; \textit{backpropagation} is a commonly used algorithm~\cite{hagan1996neural}.
ANNs architectures vary in the type of layers, activation functions, number of layers, and neurons.
Perceptron multi-layer~\cite{haykin2009neural} is a classic and commonly used ANN.
In the safety verification of ANNs, some notations are important to logically represent operations.
Hagan et al.~\cite{hagan1996neural} describe how Multi-layer Perceptrons (MLPs)
calculate a neuron output. The neuron activation potential of an input
\textit{i} = \{$i_{0}, i_{1}, i_{2}..., i_{n} $\} will be denoted as \textit{u}, which is obtained from
\begin{equation}
u = \sum_{j=0}^{n}{w_{j}\times{i}_{j}} + b.
\label{eq:anncalc}
\end{equation}
%


$\mathcal{N}$\textit{(i)} or \textit{Y} represents neuron output, that is,
$\mathcal{N}$ is the activation function of the activation potential \textit{u}.
Suppose that the activation function is \textit{sigmoid}, $\mathcal{N}$ would be described by

\begin{equation}
\mathcal{N}(u) =  \frac{\mathrm{1} }{\mathrm{1} + e^{-u}}.
\label{eq:neuronout}
\end{equation}

ReLU~\cite{li2017convergence}, Sigmoid~\cite{sibi2013analysis}, and Gaussian~\cite{sibi2013analysis}
are activation functions commonly used in MLP.
All our experimental results are obtained from MLPs with \textit{sigmoid} activation function.

However, adversarial perturbation or small noises may
lead to misclassification. In some cases, this may happen because the
network tends to not recognize the data entered due to convergence problems, e.g., the vanishing gradient
in {\it back-propagation}~\cite{bishop2006PRML}. Although some  techniques can be used to avoid
such problems (e.g., gradient regularization~\cite{RosVelez3AI2018}), safety-critical systems usually deal
with a huge volume of complex data, which may be
coded into big numbers and functions such as \textit{sigmoid} squish. Thus, a large change in the input of the
\textit{sigmoid} function will cause a small change in the output; therefore, the derivative becomes small.
For safety-critical applications, detecting where such problems may happen in a complex network is a
task that the formal verification community has been seeking to solve (e.g.,~\cite{huang2017safety});
our work aims to contribute to the solution of this research problem.

\subsection{Incremental Bounded Model Checking (BMC)}
\label{ssec:bmc}

We have used the ESBMC-GPU tool~\cite{PereiraASMMFC17,monteiro2018esbmc},
which is an extension of the Efficient SMT-Based Context-Bounded Model Checker (ESBMC)~\cite{GadelhaMMC0N18},
aimed at verifying Graphics Processing Unit (GPU)
programs written for the Compute Unified Device Architecture (CUDA) platform.
ESBMC-GPU uses an operational model (OM), i.e., an abstract representation of the standard CUDA libraries,
which conservatively approximates their semantics, to verify CUDA-based programs.

Our incremental BMC approach in ESMBC-GPU
uses an iterative technique and verifies the implementation of ANNs
for each unwind bound indefinitely or until it exhausts the time or memory limits.
We aim to either find a counterexample with up to \textit{k} loop unwinding or to fully
unwind all loops. The algorithm relies on ESBMC's
symbolic execution engine to increasingly unwind the loop after each algorithm iteration.

The approach is divided into two steps: one that tries to find property
violations and one that checks if all the loops were fully unwound. When
searching for property violation, ESBMC-GPU replaces all unwinding assertions
(i.e., assertions to check if a loop was completely unrolled) by unwinding
assumptions. Normally, this would lead to unsound behavior, however, the first
step can only find property violations, thus reporting that an unwinding assertion
failure is not a real bug. The next step is to check if all loops in the program were
fully unrolled. This is done by checking whether all the unwinding assertions are
unsatisfiable. No assertion is checked in the second step because they were already
checked in the first step for the current \textit{k} loop unwinding.

The algorithm also offers the option to change the granularity of the increment;
the default value is $1$. Note that changing the value of the increment can lead to slower
verification time and might not present the shortest counterexample possible
for the property violation. ESBMC-GPU also explicitly explores the possible interleavings
of CUDA programs (up to the given context bound), while treats each interleaving itself symbolically.
Additionally, ESBMC-GPU employs
monotonic partial order reduction~\cite{KahlonWG09} and the two-thread analysis~\cite{BettsCDQT12}
to efficiently prune the state-space exploration.
ESBMC-GPU verifies properties such as user-specified assertions, deadlocks, memory leaks,
invalid pointer dereference, array out-of-bounds, and division by zero in CUDA programs.

\section{Incremental BMC of ANNs in CUDA}
\label{sec:verification}

Our goal is to detect adversarial cases that lead ANNs to wrong results.
We have developed two verification strategies for ANNs described
using weights and bias vectors. First, we check the coverage criteria of ANNs w.r.t. a set of images.
Second, we verify safety aspects by checking adversarial cases of a given image known by an ANN.

\subsection{SMT-based Safety Verification for ANNs}
\label{ssec:method}
Our incremental verification method is based on two phases:
(1) obtain the required models from real ANN programs written in CUDA and
(2) design safety properties that ensure reliability of ANN implementations.
All models are collected from actual implementations of ANNs written in CUDA.
There exist two APIs in CUDA to support neural networks and mathematical operations:
\textit{cuBLAS}~\cite{nvidia2008cublas} and \textit{cuDNN}~\cite{chetlur2014cudnn}.
\textit{cuBLAS} provides mathematical operations such as matrix multiplication and sums;
these operations are typically used on feed-forward process of neural networks~\cite{hagan1996neural}.
\textit{cuDNN} provides deep neural networks (DNN) primitives such as tensors operations,
convolution functions, activation functions, and backward operations~\cite{hagan1996neural}.

Here we have developed operational models (cf. Section~\ref{ssec:bmc}) for the \textit{cuBLAS}
and \textit{cuDNN} libraries and integrated
them into ESBMC-GPU.\footnote{The full implementation of our OMs are available at
\url{https://github.com/LuizHenriqueSena/ESBMC-GPU/tree/master/cpp/library}}
Any ANN written in CUDA using these APIs can be verified
by our method. Additionally, any user-specified assertion can be provided
and then verified by ESBMC-GPU. Our operational models aim to perform the same steps
that the execution of the original library does but ignoring irrelevant calls (e.g., screen-printing methods),
where there exists no safety property to be checked. For the \textit{cuBLAS} and \textit{cuDNN} libraries,
we ensure that their operational models return the same results as the original APIs.
All developed modules were manually verified and exhaustively compared
with the original ones to ensure the same behavior via conformance testing~\cite{KrichenT09}.
As an example, we can see a pseudo-code of the \textit{cublasSgem} operational model
illustrated in Algorithm~\ref{euclid}, which consists of multiplying matrices $A$ and $B$
and store its result in matrix $C$.
	\begin{algorithm}
		\scriptsize
		\caption{cublasSgemm}\label{euclid}
		\begin{algorithmic}[1]
			\State $\textit{sum} \gets \text{ 0 }$
			\State $\textit{x} \gets \text{ 0 }$
			\State $\textit{y} \gets \text{ 0 }$
			\State $\textit{z} \gets \text{ 0 }$
			\For {$x < k $}
			\For {$y < i $}
			\For {$z < j $}
			\State{$sum \gets A[x][z] * B[y][x] + sum$}
			\State{$x++$}
			\EndFor
			\State{C[y][z] = sum}
			\State{$y++$}
			\EndFor
			\State{$z++$}
			\EndFor

		\end{algorithmic}
	\label{algo:1}
	\end{algorithm}

\subsection{Verification of Covering Methods}
\label{ssec:coverage}

Covering methods~\cite{sun2018testing} are based on Modified Condition/Decision
Coverage (MC/DC)~\cite{hayhurst2001practical}, which is a method applied to
ensure adequate testing for safety-critical software;
in our symbolic verification framework, MC/DC represents conditions and decisions of ANNs.
Conditions are neurons in the previously layer and decisions are neurons of the following layer.
Covering methods are used to measure how adversarial are two images w.r.t. the ANN neurons.
There exist four covering methods available in the literature~\cite{sun2018testing}:
\textit{Sign-Sign Cover} (SS Cover),
\textit{Distance-Sign Cover} (DS Cover),
\textit{Sign-Value Cover} (SV Cover),
\textit{Distance-Value Cover} (DV Cover).
Each covering method is represented as a property (assertion) within our verification framework.
In particular, each property specifies that an image set must cause that all the neurons
present in the ANN are covered by the covering method. Our algorithm uses each covering method to evaluate whether the adversarial behavior of a set of pair of images reaches a certain percentage of all containing neurons on the ANN.



In covering methods, neurons are indexed as $n_{kl}$, where $k$ represents its sort on the layer $l$, inputs are defined as $x_{1}$ and $x_{2}$. A signal-change denoted as \textit{sc} occurs when the activation potential of a certain neuron has its sign function changed by two different inputs. Sign function can be described by the equation below:
\[
sign(i) =
\begin{cases}
\text{1,} &\quad\text{if i}\ge0,\\
\text{0,} &\quad\text{otherwise.} \\
\end{cases}
\label{cas:sign}
\]


Covering  methods consist of some logical expressions:
\textit{sc(\textit{n$_{k,l}$}, \textit{x$_{1}$}, \textit{x$_{2}$}}) is denoted if the activation potential $u$ of neuron n$_{k,l}$ has its signals changed, this is, \textit{sign($u(x_{1}$))} $\neq$ \textit{sign($u(x_{2}$))}.
A value-change or \textit{vc} occurs when the activation potential of a determined neuron
represents a certain value change w.r.t. some metric $h$ and no signal change has occurred, i.e., $h(u(x1),u(x2)) = true$ and \textit{$\neg$sc(\textit{n$_{k,l}$}, \textit{x$_{1}$}, \textit{x$_{2}$}}), where $h$ can be a rate function, e.g., $\mathrm{h}(a, b)$= $\frac{a}{b} \ge d$, ($d$ represents a real number that limits the change value by $h$); a distance-change or \textit{dc} occurs when all containing neurons in a certain layer have no signal-change and its values represent some value change. We denote \textit{dc(h,k,$x_{1}$,$x_{2}$)} if \textit{$\neg$sc(\textit{n$_{k,l}$}, \textit{x$_{1}$}, \textit{x$_{2}$}}) for all neurons in layer $k$ and \textit{h(k,$x_{1}$, $x_{2}$) $\geq$ d}. As defined in value-change, function $h$ can be instantiated as any norm-based distance function and $d$ can be any real number that limits the distance function.
A neuron pair is denoted as  $\alpha$ =(\textit{n$_{k,l}$}, \textit{n$_{k+1,j}$}) and two inputs 
 are denoted as \textit{x$_{1}$} and \textit{x$_{2}$}.
The covering methods are formally described as follows:


\vspace{2mm}
\textbf{\textit{Sign Sign Cover} (\textit{SS-Cover})}:

$\neg\mathrm{sc(\textit{n$_{k,l}$}, \textit{x$_{1}$}, \textit{x$_{2}$}})$ $\forall$ \textit{p$_k,l$} $\in$ \textit{P$_k$}$\backslash$ \{i\};

$\neg\mathrm{sc(\textit{n$_{k,l}$}, \textit{x$_{1}$}, \textit{x$_{2}$}})$ $\forall$ \textit{p$_k,l$} $\in$ \textit{P$_k$}$\backslash$ \{i\};

$\mathrm{sc(\textit{n$_{k+1,l}$}, \textit{x$_{1}$}, \textit{x$_{2}$}})$;

\vspace{2mm}
\textbf{\textit{Distance Sign Cover} (\textit{DS-Cover})}:

$\mathrm{dc(\textit{h},\textit{k}, \textit{x$_{1}$}, \textit{x$_{2}$}})$;

$\mathrm{sc(\textit{n$_{k+1,l}$}, \textit{x$_{1}$}, \textit{x$_{2}$}})$;

\vspace{2mm}
\textbf{\textit{Sign Value Cover} (\textit{SV-Cover})}:

$\mathrm{sc(\textit{n$_{k,i}$}, \textit{x$_{1}$}, \textit{x$_{2}$}});$

$\neg\mathrm{sc(\textit{n$_{k,l}$}, \textit{x$_{1}$}, \textit{x$_{2}$}})$ $\forall$ \textit{p$_k,l$} $\in$ \textit{P$_k$}$\backslash$ \{i\};

$\mathrm{vc(\textit{g}, \textit{n$_{k+1,l}$}, \textit{x$_{1}$}, \textit{x$_{2}$}});$

\vspace{2mm}
\textbf{\textit{Distance Value Cover} (\textit{DV-Cover})}:

$\mathrm{dc(\textit{h},\textit{k}, \textit{x$_{1}$}, \textit{x$_{2}$}});$

$\mathrm{vc(\textit{g}, \textit{n$_{k+1,l}$}, \textit{x$_{1}$}, \textit{x$_{2}$}});$
\vspace{2mm}

\begin{figure}[H]
	\centering
	\label{fig9}
	\vspace{3ex}%
	\includegraphics[scale=0.23]{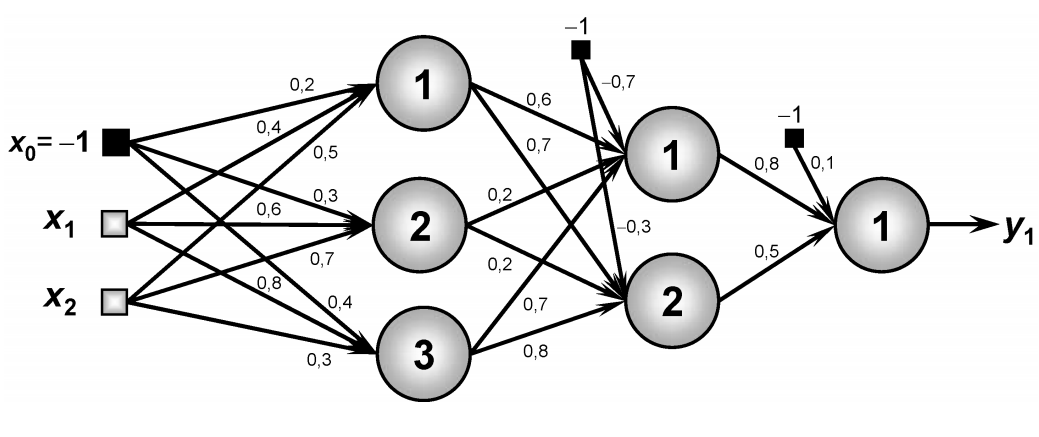}
	\caption{ANN instantiated.}
	\label{fig:netinst}
\end{figure}

Some examples of covering methods can be seen in Table~\ref{tab:cov} w.r.t. the ANN illustrated in Fig.~\ref{fig:netinst}.

\begin{table}[H]
	\centering
	\begin{tabular}{|l|l|l|l|l|l|l|l|}
		\hline
		\textbf{Ex} & \textbf{Input} & $n_{1,1}$ & $n_{2,1}$ & $n_{3,1}$ & $n_{1,2}$ & $n_{2,2}$ & $n_{1,3}$ \\ \hline
		Ex1 & (1, -3) & -1,3 & -1,80 & -0,50 & -0,79 & -1,370 & -1,417  \\ \hline
		Ex2 & (1, -1) & -0,3 & -0,40 & 0,10 & 0,51 & 0,090 & 0,353    \\ \hline
		Ex3 & (1, -1.2) & -0,4 & -0,54 & 0,04 & 0,38 & -0,056 & 0,176\\ \hline
		Ex4 & (1, -7) & -3,3 & -4,60 & -1,70s & -3,39 & -4,290 & -4,957 \\ \hline
	\end{tabular}
	\caption{ANN instantiated examples and covering methods.}
	\label{tab:cov}
\end{table}


In Table~\ref{tab:cov}, Ex1 and Ex2 instances have the pair $\alpha=\{n_{13},n_{21}\}$ covered by SS-Cover.
Here only neuron \textit{$n_{13}$} has its signal changed in layer 1. This means
that \textit{sc(\textit{n$_{1,3}$}, \textit{x$_{1}$}, \textit{x$_{2}$}}) and
$\neg$\textit{sc(\textit{n$_{1,k}$}, \textit{x$_{1}$}, \textit{x$_{2}$}}) $\forall$ \textit{p$_k,l$} $\in$ \textit{P$_k$}$\backslash$ \{1\} are \textit{true}.
Since only one neuron has its signal changed in layer $1$, any neuron in layer $2$ with its signal changed will make a pair covered by SS-Cover.
In this case, if the first neuron of layer 2 ($n_{1,2}$) has its value changed, then the pair $\alpha=\{n_{3,1},n_{1,2}\}$ is SS-Covered.

Ex2 and Ex3 are examples of DS-Cover.
In Table~\ref{tab:cov}, there exist no change values in layer $1$;
it means that for some metric \textit{h} (euclidian distance, gaussian), the two instances
of layer $1$ have a distance change and it makes \textit{dc(\textit{h},\textit{1}, \textit{x$_{1}$}, \textit{x$_{2}$}}) \textit{true}.
The distance change in layer $1$ implies a signal change of neuron \textit{$n_{2,2}$};
it makes the neuron pair $\alpha=\{n_{i,1},n_{1,2}\}$ formed by any neuron in layer $1$ covered by DS-Cover.

Ex2 and Ex3 contain a pair covered by SV-Cover. Similar to SS-Cover,
there exists only one neuron with signal change in layer $2$; it implies a value change
on neuron $n_{1,3}$, which means that the neuron pair $\alpha=\{n_{2,2},n_{1,3}\}$ is SV-Covered.
Finally, Ex1 and Ex 4 are examples of DV-Cover.
In this case, there exists no signal change in any layer of the ANN instances.
However, if there exists a metric that makes a distance change \textit{true} in layer $1$ or $2$
and if there exists another metric, which makes a value change \textit{true} in any neuron of layer $2$ or $3$,
then we have a DV-Covered pair.

The ANN properties to be checked by ESBMC-GPU generate the same literal to all four covering methods.
The neuron covered by one of all four covering methods must be equal or greater than a percentage $P$
of all containing neurons in the ANN. In particular, these properties generate literals \textit{$l_{covered\_neuron}$}
with the goal of representing the validity of the covering method w.r.t. the ANN, according to four constraints:

 For SS-Cover:
 \begin{equation}
 	l_{covered\_neuron} \Leftrightarrow (  \frac{\displaystyle \sum_{i,j}{\mathrm{ss}(n_{k,l},x_{1},x_{2})}}{N}  \geq P).
 	\label{eq:sscover}
 \end{equation}

  For DS-Cover:
 \begin{equation}
 l_{covered\_neuron} \Leftrightarrow (  \frac{\displaystyle \sum_{i,j}{\mathrm{ds}(n_{k,l},x_{1},x_{2},h)}}{N}  \geq P).
 \label{eq:dscover}
 \end{equation}

  For SV-Cover:
 \begin{equation}
 l_{covered\_neuron} \Leftrightarrow (  \frac{\displaystyle \sum_{i,j}{\mathrm{sv}(n_{k,l},x_{1},x_{2},g)}}{N}  \geq P).
 \label{eq:svcover}
 \end{equation}

  For DV-Cover:
 \begin{equation}
 l_{covered\_neuron} \Leftrightarrow (  \frac{\displaystyle \sum_{i,j}{\mathrm{ss}(n_{k,l},x_{1},x_{2},g,h)}}{N}  \geq P).
 \label{eq:dvcover}
 \end{equation}


\subsection{Verification of Adversarial Case}
\label{ssec:adversarial}

Our verification algorithm to check adversarial cases is called \texttt{checkNN}, which is
implemented in CUDA; this verification algorithm does not check the ANN code but an instance of the ANN.
This means that only weights and bias descriptions are needed for our verification process.
Let us assume an image input is represented by $I$, and \textit{m} and \textit{n} represent its size.
The dataset will be denoted as $D$; $M$ is the universe of all possible images with size \textit{m}$\times$\textit{n}. Let $\delta:\mathcal{M}^{\textit{m}\times\textit{n}}\times \mathcal{M}^{\textit{m}\times\textit{n}} \rightarrow \mathbb{R}$ be an euclidian distance operator defined as follows:

\begin{equation}
\delta(P, Q) = \sqrt{\sum_{i=0}^{n}{(p_{i} - q_{i})^2}}.
\label{eq:eucliddist}
\end{equation}

Suppose that any image can be casted to a vector and its values are normalized.
$P$ = ($p_{1}, p_{2}, p_{3}, \ldots, p_{n}$) and $Q = (q_{1}, q_{2}, q_{3}, \ldots, q_{n}) \in
\mathcal{R}^{N}$. Suppose that $P$ and $Q$ are 5$\times$5 images, i.e., both vectors length are 25 and they are represented by the images ``A'' and ``O'', respectively, as illustrated below:
\begin{figure}[H]
	\centering
	\begin{subfigure}[b]{0.1\linewidth}
		\centering
		\includegraphics[width=\linewidth]{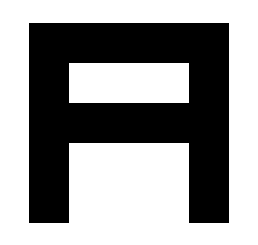}
		\label{figA:A}
	\end{subfigure}
	\begin{subfigure}[b]{0.1\linewidth}
		\centering
		\includegraphics[width=\linewidth]{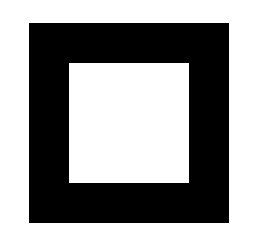}
		\label{figO:O}
	\end{subfigure}
	\caption{Vocalic ``A'' and ``O''.}\label{fig:vocalicao}
\end{figure}

If we apply these two images in Fig.~\ref{fig:vocalicao} to the euclidian distance equation described by Eq.~\eqref{eq:eucliddist}, $\delta$ will return $2.449$. 
%
%
Denoting ${I}^{d}$ as an image that belongs to the dataset, all sets can be represented as follows: 

\begin{equation}
\centerline{${I}^{d} \in \mathcal{D}^{\textit{m}\times\textit{n}}$, ${I}$ $\in$ $\mathbb{M}^{\textit{m}\times\textit{n}}$, $\mathcal{D}^{\textit{m}\times\textit{n}}$ $\subseteq$ $\mathbb{M}^{\textit{m}\times\textit{n}}$.}
\end{equation}

Our verification algorithm \texttt{checkNN} requires three inputs:
the first input is an ANN with its weights and bias descriptions;
the second one is the image to be checked; and the
third one is a parameter that limits the proximity of the adversarial cases.
Our safety property is specified using two intrinsic operators supported by ESBMC-GPU:
\texttt{assume} and \texttt{assert}. ESBMC-GPU assumes that a function call
\texttt{assume(expression)} has the following meaning:
if ``expression'' is evaluated to ``0'', then the function loops forever;
otherwise the function returns (no side effects). In particular,
the desired image to be checked can be represented by
an \texttt{assume}, which is formally described as the restriction $R$: 

\begin{equation}
R = I \in \mathbb{M}^{mn} \mid \delta ({I}^{d},\mathcal{I}) \leq b \}.
\label{eq:restriction}
\end{equation}

Eq.~\eqref{eq:restriction} denotes that $\mathcal{I}^{d} \in \mathcal{M}^{mn}$, which
will be compared with a non-deterministic image $\mathcal{I}$
until the euclidian distance $\delta (\mathcal{I}^{d},\mathcal{I})$ described by Eq.~\eqref{eq:eucliddist}
is less than or equal to the parameter $b$.
After a non-deterministic image $\mathcal{I}$ is obtained,
a safety property is checked by ESBMC-GPU, which is represented
by an \texttt{assert} statement formally described as

\begin{equation}
\centerline{$\mathcal{Y}^{d}$ = \{$N(\mathcal{I})$ $ ,$ $\forall$ $\mathcal{I} \in \mathcal{R}\}$.}
\label{eq:assert}
\end{equation}



In Eq.~\eqref{eq:assert}, $\mathcal{Y}^{d}$ represents any ANN's output mapped by the input $\mathcal{I}^{d}$ on the given dataset.
The function $N$ represents the ANN function $\mathcal{R}^{m} \rightarrow \mathcal{R}^{n}$.
The property is taken as the negative of Eq.~\eqref{eq:assert}.
If the output obtained by function $N$ and the non-deterministic image
is different from the mapped output, then the property is violated and a counterexample is produced.
The property described by Eq.~\eqref{eq:assert} generates the literal \textit{$l_{image\_misclassified}$}.
A classification is obtained from the output values represented
by the neurons of the last layer. A reference value is denoted by the variable $V$,
a desired classification is denoted by variable $D$, which represents
the neuron position of the output layer and \textit{i} represents any other
neuron position different than $D$. The literal \textit{$l_{image\_misclassified}$}
represents the validity of the original image classification, according to the constraint:

\begin{equation}
l_{image\_misclassified} \Leftrightarrow (n_{D,3} < V) \land (n_{i,3} > V).
\end{equation}

\section{Experimental Evaluation}
\label{sec:exp}


\subsection{Description of the Benchmarks}
\label{ssec:benchmarks}

\begin{figure}[!h]
	\centering
	\begin{subfigure}[b]{0.1\linewidth}
		\centering
		\includegraphics[width=\linewidth]{imagema.png}
		\label{fig:A}
	\end{subfigure}
	\begin{subfigure}[b]{0.1\linewidth}
		\centering
		\includegraphics[width=\linewidth]{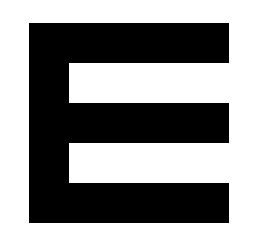}
		\label{fig:E}
	\end{subfigure}
	\begin{subfigure}[b]{0.1\linewidth}
		\centering
\includegraphics[width=\linewidth]{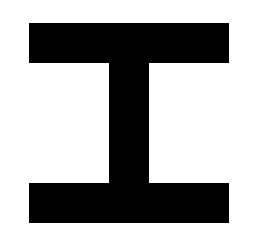}
\label{fig:I}
	\end{subfigure}
\begin{subfigure}[b]{0.1\linewidth}
		\centering
\includegraphics[width=\linewidth]{imagemo.png}
\label{fig:O}
\end{subfigure}
\begin{subfigure}[b]{0.1\linewidth}
		\centering
\includegraphics[width=\linewidth]{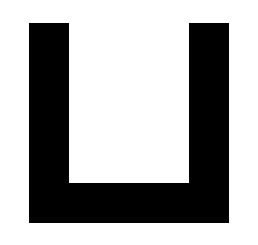}
\label{fig:U}
\end{subfigure}
	\caption{Vocalic images in benchmarks.}\label{fig:vowels}
\end{figure}

Our evaluation employs a new character pattern recognition benchmark,
which was developed by our research group with the goal of performing
conformance testing~\cite{KrichenT09} over our \textit{cuBLAS} and \textit{cuDNN} operational models
(cf. Section~\ref{ssec:method})\footnote{Available at \url{https://github.com/LuizHenriqueSena/ESBMC-GPU/tree/master/cpp/library}}
In particular, each experiment uses an ANN that solves the problem of
vocalic pattern recognition in $5x5$ images as input. The ANN was trained by
the back-propagation algorithm~\cite{bishop2006PRML}. Our dataset is composed by $100$ correct vocalics
with noise images and $100$ non vocalic images. All vocalics are illustrated in Fig.~\ref{fig:vowels}.

We conducted the experimental evaluation on a 8-core 3.40 GHz
Intel Core i7 with 24 GB of RAM and Linux OS.
We use CUDA v9.0, cuDNN v5.0, cuBLAS v10.1, and
ESBMC-GPU v2.0.\footnote{Available at \url{http://gpu.esbmc.org/}}
All presented execution times are actually CPU times, i.e.,
only the elapsed time periods spent in the allocated CPUs,
which was measured with the times system call (POSIX system).
All adversarial cases were obtained by executing the
following command, which is wrapped in a shell script to
iteratively unwind the program:
 \texttt{esbmc-gpu verifynn.c -I  <path-to-OM> --force-malloc-success --no-div-by-zero-check --no-pointer-check --no-bounds-check --incremental-bmc}.


\subsubsection{Availability of Data and Tools}
\label{ssec:AvailabilityofDataandTools}

All benchmarks, tools, and results associated with the current evaluation,
are available for downloading at \url{https://github.com/ssvlab/ssvlab.github.io/tree/master/gpu/benchmarks}.

\subsection{Objectives}
\label{ssec:desc_exp}

Using the benchmarks given in Section~\ref{ssec:benchmarks},
our evaluation has the following two experimental goals:

\begin{enumerate}

\item[EG1] \textbf{(Covering Methods)} Evaluate the performance and correctness of our symbolic verification algorithms to check all four covering methods (cf.~Section~\ref{ssec:coverage}).

\item[EG2] \textbf{(Adversarial Cases)} Evaluate the performance and correctness of our verification algorithm \texttt{checkNN} (cf. Section~\ref{ssec:adversarial}) to verify adversarial cases obtained from changing input images and parameter proximity.

\end{enumerate}

\subsection{Results}
\label{ssec:results}

\subsubsection{Covering Methods}
\label{exp:CoveringMethods }

in covering methods experiments, ESBMC-GPU was able to correctly verify all four methods:
SS-Cover, DS-Cover, SV-Cover, and DV-Cover. The verification time of all four covering methods
did not take longer than a few seconds for checking how adversarial are two images w.r.t. the ANN neurons as described
by Equations~\eqref{eq:sscover},~\eqref{eq:dscover},~\eqref{eq:svcover}, and~\eqref{eq:dvcover} in Section~\ref{ssec:coverage}.
The fast verification was expected since all covering methods
have only deterministic inputs. These results successfully answer \textbf{EG1}: ESBMC-GPU
achieves reasonable performance to correctly verify all covering methods.

The verified properties imply that $80$\% of all neurons on the ANN must be covered by the set of images.
In particular, the set of images employed was the dataset used during the training phase;
the tool output of all benchmarks correctly returned that covered neurons were not greater than $80$\%.
As a result, the dataset was unable to provide $80$\% of neuron coverage by any covering method.
With respect to this property, the average execution time of all four covering methods, when applied to a set of $200$ images,
is around $20$ minutes; additionally, ESBMC-GPU does not generate test cases based on covering methods yet, which we leave for future work.

\begin{figure}[!h]
	\centering
	\begin{subfigure}[b]{0.1\linewidth}
		\centering
		\includegraphics[width=\linewidth]{imagemu.png}
		\label{fig:uori}
	\end{subfigure}
	\begin{subfigure}[b]{0.1\linewidth}
		\centering
		\includegraphics[width=\linewidth]{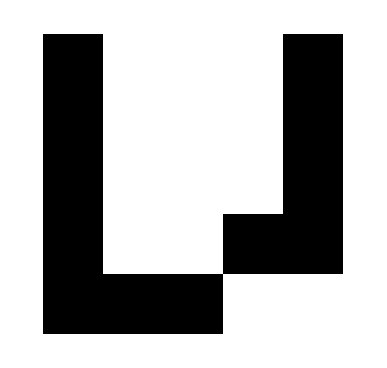}
		\label{fig:unoised}
	\end{subfigure}
	\caption{Vocalics U and U with some noise.}\label{fig:vocalicuunoised}
\end{figure}

The two generated ANN instances of the two inputs ``U'' and noise ``U'' illustrated in Fig.~\ref{fig:vocalicuunoised} can be seen in Table~\ref{tab:instancesU}.

\begin{table}[!h]
	\centering
	\begin{tabular}{|c|c|c|}
		\hline
		\textbf{Neuron} &\textbf{Image "U"} & \textbf{Noise "U"} \\ \hline
		$n_{1,1}$ &-1.885322 &  4.619613  \\ \hline
		$n_{2,1}$ &8.775419 &  9.796190   \\ \hline
		$n_{3,1}$ &2.959348 &  5.743809   \\ \hline
		$n_{4,1}$ &10.424796 &  4.046428   \\ \hline
		$n_{5,1}$ &8.172012 &  14.466885  \\ \hline
		$n_{1,2}$ &-3.863095 &  -9.308636   \\ \hline
		$n_{2,2}$ &5.328067 &  5.263461   \\ \hline
		$n_{3,2}$ &-3.770385 &  -5.705760   \\ \hline
		$n_{4,2}$ &0.574238 &  -2.029373   \\ \hline
		$n_{1,3}$ &-6.707186 &  -7.149290   \\ \hline
		$n_{2,3}$ &-15.815082 &  -17.246468   \\ \hline
		$n_{3,3}$ &-10.060704 &  -13.074245   \\ \hline
		$n_{4,3}$ &-9.688183 &  -4.868999   \\ \hline
		$n_{5,3}$ &-0.555885 &  3.355738   \\ \hline

	\end{tabular}
	\caption{ANN instances of two inputs.}
	\label{tab:instancesU}
\end{table}

According to the SS-Cover equation described in Section~\ref{sec:verification}
and the literal described by Eq.~\eqref{eq:sscover}, we have two pairs of neuron:
$\alpha1=\{n_{1,1}, n_{4,2}\}$ and $\alpha2=\{n_{4,2}, n_{5,3}\}$
SS-Covered by the two images. Only $3$ of $14$ neurons are SS-Covered, that is,
only $21$\% of the neurons are covered by the covering method.
The literal described by Eq.~\eqref{eq:sscover} specifies that the neuron
coverage must be greater than a percentage $P$.
All benchmarks were run with the value of $P$ set to $80$\%,
which means that for these two inputs the property fails.

\subsubsection{Adversarial Cases}
\label{exp:AdversarialCases}

in adversarial cases, we executed $14$ experiments, where the average verification time was approx. $19$ hours.
Table~\ref{tab:time} shows the verification time and parameter values of our experiments.
Here, \textit{benchmark} is the experiment identifier,
\textit{image} represents the desired image, where we intend to have the classification verified,
$\gamma$ represents a limit of proximity for each benchmark, and
\textit{verification time} is the execution time taken by ESBMC-GPU measured in hours using
times system call (POSIX system).

\begin{table}[H]
	\centering
	\begin{tabular}{|c|c|c|c|}
		\hline
		\textbf{Benchmark} &\textbf{Image} & \textbf{$\gamma$} & \textbf{Verification Time (hours)}\\ \hline
		Ex1 &Vocalic O & 0.5 & 1 \\ \hline 
		Ex2 &Vocalic O & 1.5 & 4 \\ \hline 
		Ex3 &Vocalic O & 2.5 & 8 \\ \hline 
		Ex4 &Vocalic O & 3.5 & 6 \\ \hline 
		Ex5 &Vocalic E & 0.5 & 25 \\ \hline 
		Ex6 &Vocalic E & 0.7 & 25 \\ \hline 
		Ex7 &Vocalic E & 1.5 & 14 \\ \hline 
		Ex8 &Vocalic E & 3.0 & 12 \\ \hline 
		Ex9 &Vocalic U & 0.3 & 6 \\ \hline 
		Ex10 &Vocalic U & 0.5 & 5 \\ \hline 
		Ex11 &Vocalic U & 1.0 & 20 \\ \hline 
		Ex12 &Vocalic U & 1.5 & 19 \\ \hline 
		Ex13 &Vocalic A & 1.0 & 63 \\ \hline 
		Ex14 &Vocalic A & 1.5 & 63 \\ \hline 

	\end{tabular}
	\caption{Verification time and proximity parameter relation.}
	\label{tab:time}
\end{table}

Incremental BMC approach has not been so efficient w.r.t. verification time since we
verify the actual implementation of ANNs in CUDA using the logics QF\_AUFBV from the SMT standard~\cite{Barrett10c},
The QF\_AUFBV logic represents quantifier-free formulae that are built over
bit-vectors and arrays with free sort and function symbols, but with the
restriction that all array terms have the following structure
(array (bit-vector $i$[$w_1$]) (bit-vector $v$[$w_2$])), where $i$ is the index
with bit-width $w_1$ and $v$ is the value with bit-width $w_2$.

Although our ANN benchmarks contain $3$ layers, ESBMC-GPU took longer than other existing tools,
which have previously reported experimental results with larger ANNs~\cite{huang2017safety}.
In particular, DLV (Deep Learning Verification)~\cite{huang2017safety} has obtained adversarial cases
of ANNs with $12$ layers, ranging from a few seconds to $20$ minutes.
In order to check refinement by layer, DLV uses the theory of linear real arithmetic with
existential and universal quantifiers, and for verification within a layer (0-variation), DLV uses the
same theory but without universal quantification.

\begin{figure}[!h]
	\centering
	\captionsetup[subfigure]{labelformat=empty}
	\begin{subfigure}[t]{0.45\linewidth}
		\centering
		\begin{subfigure}[t]{0.45\linewidth}
			\centering
			\includegraphics[height=0.8in]{imagemo.png}
		\end{subfigure}%
		~
		\begin{subfigure}[t]{0.45\linewidth}
			\centering
			\includegraphics[height=0.8in]{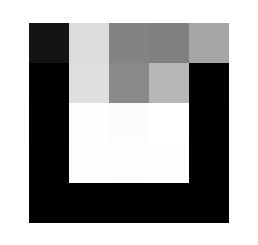}
		\end{subfigure}
		\caption{(a) Vocalic "O" misclassified as "U".$\gamma$=0.5}
	\end{subfigure}
	\begin{subfigure}[t]{0.45\linewidth}
		\centering
		\begin{subfigure}[t]{0.45\linewidth}
			\centering
			\includegraphics[height=0.8in]{imagemo.png}
		\end{subfigure}%
		~
		\begin{subfigure}[t]{0.45\linewidth}
			\centering
			\includegraphics[height=0.8in]{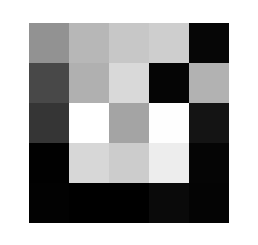}
		\end{subfigure}
		\caption{(b) Vocalic "O" misclassified as "U".$\gamma$=0.7}
	\end{subfigure}
	\begin{subfigure}[t]{0.45\linewidth}
		\centering
		\begin{subfigure}[t]{0.45\linewidth}
			\centering
			\includegraphics[height=0.8in]{imagemo.png}
		\end{subfigure}%
		~
		\begin{subfigure}[t]{0.45\linewidth}
			\centering
			\includegraphics[height=0.8in]{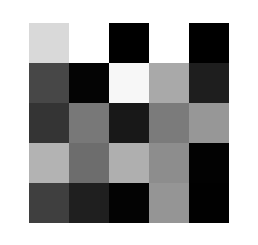}
		\end{subfigure}
		\caption{(c) Vocalic "O" misclassified as "U".$\gamma$=1.5}
	\end{subfigure}
	\begin{subfigure}[t]{0.45\linewidth}
		\centering
		\begin{subfigure}[t]{0.45\linewidth}
			\centering
			\includegraphics[height=0.8in]{imagemo.png}
		\end{subfigure}%
		~
		\begin{subfigure}[t]{0.45\linewidth}
			\centering
			\includegraphics[height=0.8in]{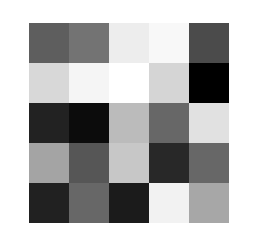}
		\end{subfigure}
		\caption{(d)Vocalic "O" misclassified as "U".$\gamma$=3.0}
	\end{subfigure}
%
%
	\begin{subfigure}[t]{0.45\linewidth}
		\centering
		\begin{subfigure}[t]{0.45\linewidth}
			\centering
			\includegraphics[height=0.8in]{imageme.png}
		\end{subfigure}%
		~
		\begin{subfigure}[t]{0.45\linewidth}
			\centering
			\includegraphics[height=0.8in]{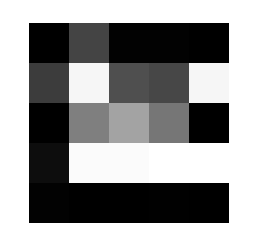}
		\end{subfigure}
		\caption{(e) Vocalic "E" misclassified as "O".$\gamma$=0.5}
	\end{subfigure}
	\begin{subfigure}[t]{0.45\linewidth}
		\centering
		\begin{subfigure}[t]{0.45\linewidth}
			\centering
			\includegraphics[height=0.8in]{imageme.png}
		\end{subfigure}%
		~
		\begin{subfigure}[t]{0.45\linewidth}
			\centering
			\includegraphics[height=0.8in]{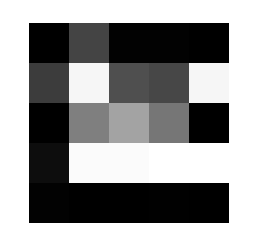}
		\end{subfigure}
		\caption{(f) Vocalic "E" misclassified as "O".$\gamma$=0.7}
	\end{subfigure}
	\begin{subfigure}[t]{0.45\linewidth}
		\centering
		\begin{subfigure}[t]{0.45\linewidth}
			\centering
			\includegraphics[height=0.8in]{imageme.png}
		\end{subfigure}%
		~
		\begin{subfigure}[t]{0.45\linewidth}
			\centering
			\includegraphics[height=0.8in]{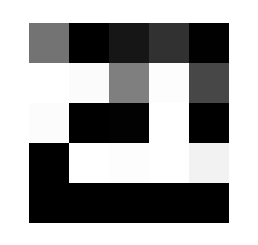}
		\end{subfigure}
		\caption{(g) Vocalic "E" misclassified as "O".$\gamma$=1.5}
	\end{subfigure}
	\begin{subfigure}[t]{0.45\linewidth}
		\centering
		\begin{subfigure}[t]{0.45\linewidth}
			\centering
			\includegraphics[height=0.8in]{imageme.png}
		\end{subfigure}%
		~
		\begin{subfigure}[t]{0.45\linewidth}
			\centering
			\includegraphics[height=0.8in]{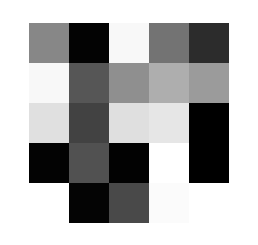}
		\end{subfigure}
		\caption{(h)Vocalic "E" misclassified as "O".$\gamma$=3.0}
	\end{subfigure}
	\begin{subfigure}[t]{0.45\linewidth}
	\centering
	\begin{subfigure}[t]{0.45\linewidth}
		\centering
		\includegraphics[height=0.8in]{imagemu.png}
	\end{subfigure}%
	~
	\begin{subfigure}[t]{0.45\linewidth}
		\centering
		\includegraphics[height=0.8in]{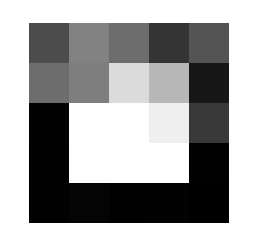}
	\end{subfigure}
	\caption{(i)Vocalic "U" misclassified as "O".$\gamma$=0.3}
	\end{subfigure}
	\begin{subfigure}[t]{0.45\linewidth}
	\centering
	\begin{subfigure}[t]{0.45\linewidth}
		\centering
		\includegraphics[height=0.8in]{imagemu.png}
	\end{subfigure}%
	~
	\begin{subfigure}[t]{0.45\linewidth}
		\centering
		\includegraphics[height=0.8in]{iuo3.png}
	\end{subfigure}
	\caption{(j)Vocalic "U" misclassified as "O".$\gamma$=0.5}
\end{subfigure}
	\begin{subfigure}[t]{0.45\linewidth}
	\centering
	\begin{subfigure}[t]{0.45\linewidth}
		\centering
		\includegraphics[height=0.8in]{imagemu.png}
	\end{subfigure}%
	~
	\begin{subfigure}[t]{0.45\linewidth}
		\centering
		\includegraphics[height=0.8in]{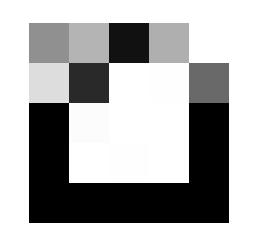}
	\end{subfigure}
	\caption{(l)Vocalic "U" misclassified as "O".$\gamma$=1.0}
\end{subfigure}
	\begin{subfigure}[t]{0.45\linewidth}
	\centering
	\begin{subfigure}[t]{0.45\linewidth}
		\centering
		\includegraphics[height=0.8in]{imagemu.png}
	\end{subfigure}%
	~
	\begin{subfigure}[t]{0.45\linewidth}
		\centering
		\includegraphics[height=0.8in]{iuo10.png}
	\end{subfigure}
	\caption{(k)Vocalic "U" misclassified as "O".$\gamma$=1.5}
\end{subfigure}
%
	\caption{Generated adversarial cases.}
	\label{res:adver}
\end{figure}

In prior work, Cordeiro et al.~\cite{CordeiroFM12}
reported that, although verification conditions are solved faster using the theory of linear real arithmetic
(since the result of the analysis is independent from the actual binary representation),
the theory of bit-vector allows the encoding of bit-level operators more accurately, which is inherently present
in the implementation of ANNs. These results partially answer \textbf{EG2}: ESBMC-GPU is able to correctly produce
adversarial cases, which are confirmed by graphical inspection in MATLAB, but our
verification time is high due to the bit-accurate precision
of our verification model. Some of the adversarial cases produced by ESBMC-GPU are shown in Fig.~\ref{res:adver}.

Unfortunately, a direct performance comparison using the same ANN with
DLV was not possible due to some compatibility issues.
Note that DeepConcolic~\cite{sun2018testing} and DLV~\cite{huang2017safety}
perform verification and testing only on ANNs trained by classics dataset as MNIST;
both tools do not support the verification and testing on general ANNs.
For future work, we will adapt our tool to support the same models as the ones handled
by DeepConcolic and DLV.

\subsection{Threats to validity}
\label{sec:ThreatstoValidity}

Although all adversarial cases obtained from the benchmarks are real,
which we confirmed by executing our validation scripts in MATLAB, our algorithm uses a lookup
table to compute the \textit{sigmoid} activation function;
even with a good resolution, a lookup table will always contain more value errors
than a computed function, which could potentially produce a false adversarial case.
Additionally, our new character pattern recognition benchmark is probably
an easy ANN to find adversarial cases, since the ANN training
was performed by a small dataset, which contains only $200$ images. This kind of training probably
generates ANN with a low-level of safety classification. Another identified threat is that our
verification framework is not flexible enough; any change to ANN layer number
might not be an easy task since our verification algorithms based on incremental BMC
were developed using \textit{cuBLAS} and \textit{cuDNN} primitives;
these primitives are not so flexible to perform such changes.

\section{Related work}
\label{sec:related_work}

Our ultimate goal is to formally ensure safety for applications that are based on Artificial Intelligence (AI),
as described by Amodei et al.~\cite{amodei2016concrete}. In particular, the potential impact of intelligent
systems performing tasks in society and how safety guarantees are necessary to prevent damages
are the main problem of safety in ANNs.

Sun et al.~\cite{sun2018testing} and Huang et al.~\cite{huang2017safety} have shown how weak ANNs can
be if small noises are present in their inputs. They described and evaluated testing and verification approaches based on
covering methods and images proximity~\cite{sun2018testing} and how adversarial cases are obtained~\cite{huang2017safety}.
In particular, our study resembles that of Huang et al. and Su et al.~\cite{huang2017safety,sun2018testing} to obtain adversarial cases.
Here, if any property is violated, then a counterexample is provided; in cases of safety properties,
adversarial examples will be generated via counterexample using ESBMC-GPU.
In contrast to Huang et al.~\cite{huang2017safety}, we do not focus on generating noise in specific regions,
but in every image pixel. Our approach in images proximity is influenced by Sun et al.~\cite{sun2018testing} but
we use incremental BMC instead of concolic testing as our verification engine.
Our symbolic verification method checks safety properties on non-deterministic images
with a certain distance of a given image; both image and distance are determined by the user.
Gopinath et al.~\cite{abs-1807-10439} also describe an approach to validate ANNs using symbolic execution
by translating a NN into an imperative program. By contrast, we consider the actual implementation
of ANN in CUDA and apply incremental BMC using off-the-shelf SMT solvers.

Gopinath et al.~\cite{DBLP:journals/corr/abs-1904-13215} presented formal
techniques to extract invariants from the decision logic of ANNs.
These invariants represent pre- and post-conditions, which hold when transformations
of a certain type are applied to ANNs. The authors have proposed two techniques.
The first one is called iterative relaxation of decision patterns, which uses Reluplex
as the decision procedures~\cite{katz2017reluplex}. The second one is called decision-tree based invariant generation, which
resembles covering methods~\cite{sun2018testing}. Robustness and explainability are the core properties
of this study. Applying those properties to ANNs have shown impressive experimental results. Explainability
showed an important property to evaluate safety in ANNs; the core idea is to obtain
explanations of why the adversarial case happened by observing the pattern activation behavior
of a subset of neurons described by the given invariant.

Gopinath et al.~\cite{DBLP:journals/corr/abs-1710-00486} also proposed a novel approach for automatically
identifying safe regions of inputs w.r.t. some labels. The core idea is to identify safe regions w.r.t. labeled targets,
i.e., providing a specific safety guarantee that a robust region is robust enough against adversarial perturbations w.r.t.
to a target label. As the notion of safety robustness in ANNs is a strong term for many ANNs, the target robustness
is the main property. The technique works with clustering and verification. Clustering technique is used
to split the dataset into a subset of inputs with the same labels, then each cluster is verified by Reluplex~\cite{katz2017reluplex}
to provide the safety region w.r.t. the target label. The tool proposed is called DeepSafe, which is evaluated on trained ANNs
by the dataset MNIST and ACAS XU.

In addition to ESBMC-GPU, there exist other tools able to verify CUDA programs
and each one of them uses its approach and targets specific property violations.
However, given the current knowledge in software verification, ESBMC-GPU is the first verifier
to check for adversarial cases and coverage methods in ANNs implemented in CUDA. For instance, GPUVerify~\cite{BettsCDQT12}
is based on synchronous, delayed visibility semantics, which focuses on detecting data race and barrier divergence,
while reducing kernel verification procedures for the analysis of sequential programs.
GPU+KLEE (GKLEE)~\cite{LiLSGGR12}, in turn, is a concrete and symbolic execution tool,
which considers both {\it kernels} and {\it main} functions, while checking deadlocks,
memory coalescing, data race, warp divergence, and compilation level issues.
Also, Concurrency Intermediate Verification Language (CIVL)~\cite{ZhengRLDS15},
a framework for static analysis and concurrent program verification,
uses abstract syntax tree and partial order reduction to detect user-specified assertions,
deadlocks, memory leaks, invalid pointer dereference, array out-of-bounds, and division by zero.

Our approach implemented on top of ESBMC-GPU has some similarities with
other techniques described here, e.g., covering methods proposed by Sun et al.~\cite{sun2018testing},
model checking to solve adversarial cases proposed by Huang et al.\cite{huang2017safety}.
However, the main contribution is our requirements and how we handle the actual implementations of ANNs.
To run our proposed safety verification, only the ANNs with weights and bias descriptors
and the desired input of the dataset is required. For tools such as DeepConcolic~\cite{sun2018testing} and DLV~\cite{huang2017safety},
obtaining adversarial cases or safety guarantees for different ANNs is not an easy task due to the focus given
to the famous datasets as MNIST~\cite{mnist} or CIFAR-10~\cite{cifar10} during the tool development.
In our proposed approach, there exists no need for providing specific datasets,
but only the desired dataset sample to be verified. Besides these requirements,
it is necessary for the user to know how cuDNN~\cite{chetlur2014cudnn} deals with ANNs.

\section{Conclusions}
\label{sec:conclusion}

We have described and evaluated two approaches for verifying ANNs:
one to check for adversarial cases and another one to check for
coverage methods in MLP.
In particular, our verification method was able to find adversarial cases
for different input images and proximity parameter values.
Despite a high verification time in some benchmarks due to bit-accurate verification,
the average time is reasonable since our approach exhaustively verifies
all possible adversarial cases. In other approaches,
obtaining adversarial cases of even bigger ANNs tend to be faster,
but noises are not fully explored.
%
%

Our verification of covering methods was able to correctly verify our dataset and
has shown to be effective during our experiments.
All OMs developed here are integrated into ESBMC-GPU
and can be used to verify other programs that use the \textit{cuBLAS} and \textit{cuDNN} APIs,
while other existing approaches (e.g., DeepConcolic and DLV)
perform verification and testing only on ANNs trained by classics dataset as MNIST, i.e.,
they do not offer much flexibility to perform verification and testing on general ANNs.

Future work aims to implement further techniques (e.g., invariant inference~\cite{GadelhaMCN19})
to prune the state space exploration, by taking into account design aspects of ANNs.
We will also improve our training dataset by considering the counterexample produced by our verification framework.
Lastly, we will investigate fault localization and repair techniques~\cite{AlvesCF17} to make the ANN implementation
robust against small noises that are present in the ANN inputs.





\end{document}